\documentclass{jpp}
\usepackage{graphicx}
\usepackage{color}

\title{Gyrokinetic continuum simulation of turbulence in a straight open-field-line plasma}

\author{E. L. Shi\aff{1}
  \corresp{\email{eshi@princeton.edu}},
  G. W. Hammett\aff{2},
  T. Stoltzfus-Dueck\aff{1,3}
  \and A. Hakim\aff{2} }

\affiliation{\aff{1}Department of Astrophysical Sciences,
  Princeton University, Princeton, NJ 08540, USA
\aff{2}Princeton Plasma Physics Laboratory, Princeton, NJ
  08543-0451, USA
\aff{3}Max-Planck-Princeton Center for Plasma Physics,
  Princeton University, Princeton, NJ 08540, USA}

\begin{document}

\maketitle

\begin{abstract}
3D2V continuum gyrokinetic simulations of electrostatic plasma turbulence
in a straight, open-field-line geometry have been performed using the
full-$f$ discontinuous-Galerkin code Gkeyll.
These simulations include the basic elements of a fusion-device scrape-off layer:
localized sources to model plasma outflow from the core,
cross-field turbulent transport, parallel flow along magnetic field lines,
and parallel losses at the limiter or divertor with sheath model boundary conditions.
The set of sheath boundary conditions used in the model allows currents to flow through the walls.
In addition to details of the numerical approach, results from numerical simulations of turbulence in
the Large Plasma Device (LAPD), a linear device featuring straight magnetic field lines, are presented.
\end{abstract}

\section{Introduction}

The scrape-off layer (SOL) of a tokamak plasma is a region at the outermost edge where the plasma flows unconstrained along 
open magnetic field lines that intersect the material wall \citep{Stangeby2000,SoltzfusDueck2009}.
The SOL plasma sets the boundary conditions on the core confined plasma, so the ability to influence
the plasma behaviour in this region and in the pedestal is key to improving overall reactor performance
\citep{Zweben2007,Kotschenreuther1995,Dimits2000,Kinsey2011,Shimada2007}.
On a basic level, plasma dynamics in the SOL involve plasma outflow from the core,
cross-field turbulent transport, and parallel losses at the divertor or limiter plates \citep{Mosetto2014},
where plasma-surface interactions such as recycling and impurity influx can occur.
The balance of these processes is believed to set the SOL width, which affects the location and strength of
heat loads on plasma-facing components \citep{Eich2013}.
The ability to operate future tokamaks like ITER and DEMO with high fusion gain without the significant erosion and melting of
plasma-facing components is a major challenge that necessitates a thorough understanding of SOL turbulence.

One of the most common approaches to model the SOL is to solve simplified transport equations based on the Braginskii fluid equations.
These codes use approximate sheath boundary conditions in the parallel direction and do not capture plasma turbulence,
requiring the use of ad-hoc diffusion terms to model the turbulent transport across the magnetic field \citep{Schneider2006,Rognlien1994}.
Edge turbulence codes solving 3D drift-reduced Braginskii fluid equations have also been developed \citep{Ricci2008,Dudson2009,Scott1997,Xu1998}.
While Braginskii-fluid turbulence codes are relatively fast and have led to new insights into edge turbulence,
they omit kinetic effects by approximating the plasma as highly collisional, an assumption that is typically violated
in the tokamak SOL.
Some codes that can include the SOL \citep{Ribeiro2005, Xu2013}
have implemented gyrofluid models, which are more general and address these issues
to some extent by incorporating finite-Larmor-radius effects and a Landau-damping model
\citep{Dorland1993,Snyder1997}, but they can only approximate certain nonlinear effects,
such as the treatment of energetic tails, which are important for sheath physics.
For these reasons, there are efforts to develop first-principles gyrokinetic codes for edge turbulence simulation \citep{Chang2009,Dorf2016,Korpilo2016}.
Unlike Braginskii fluid approaches, gyrokinetic approaches use equations that are valid across a wide range of collisionality regimes,
even if the collisional mean free path is not small compared to the parallel scale length
or if the ion drift orbit excursions are not small compared to radial gradient length scales \citep{Dorf2016}.
Gyrokinetic simulations, however, are much more computationally expensive than their fluid counterparts,
so both approaches can be useful.

Continuum methods are Eulerian approaches to solve a kinetic equation (e.g. the 5D gyrokinetic equation)
by discretizing the equation on a phase-space mesh.
The other main class of algorithms used for plasma simulation is the particle-in-cell (PIC) method,
which is essentially a Monte Carlo sampling technique
that uses macroparticles to integrate along the characteristic phase-space trajectories of many gyrocentres without
the need for a velocity-space grid \citep{Krommes2012}.
Each method has advantages, disadvantages, and challenges,
and it is important to explore both approaches as independent cross-checks against each other and to guide the development
of future gyrokinetic edge turbulence codes.
Gyrokinetic PIC codes that include a scrape-off-layer region have been developed with various capabilities and are being extended \citep{Chang2009,Korpilo2016},
while gyrokinetic continuum codes for edge simulation are less mature.

The edge region is challenging to simulate for a number of reasons, including the need to
handle large-amplitude fluctuations while avoiding negative overshoots,
open and closed field lines with a separatrix and X-point
(which can cause difficulties with coordinates),
fully electromagnetic fluctuations near the beta limit,
a wide range of space and time scales, a wide range of collisionality regimes,
sheath boundary conditions, plasma-wall interactions, atomic physics,
and the existence of very high frequency $\omega_H$ modes \citep{Lee1987,Belli2005}
or sheath-interaction modes that one does not want to artificially excite.
Sophisticated gyrokinetic codes for the core region of tokamaks have been
developed and are highly successful, but major extensions to them or new
codes are required to handle the additional challenges of the edge region.

Many of the existing core gyrokinetic codes assume small amplitude fluctuations.
Many of them use spectral techniques in some directions, which can have problems
with Gibbs phenomena that give negative overshoots in the solution.
Most algorithms used in magnetic fusion research are designed for
cases where viscous or dissipative scales are fully resolved and do not use limiters,
and thus can have problems with small negative oscillations.
Negative densities may cause various unphysical problems in the solution
(for example, a negative density in the tail of the electron distribution
function can reverse the slope of the sheath current versus sheath potential
relation).
Some finite-difference algorithms make it easier to calculate derivatives
across the separatrix with field-aligned coordinates, but may have problems
with particle conservation, and small imbalances in electron and ion gyrocentre
densities may drive large electric fields.

Gkeyll is a plasma simulation code that implements several fluid and kinetic models using
a variety of grid-based numerical algorithms.
Recently, Gkeyll has been used for fluid studies of magnetic reconnection \citep{Wang2015,Ng2015}
and kinetic and multi-fluid sheath modelling \citep{Cagas2016}.
The work presented here focuses on our efforts to implement gyrokinetic continuum algorithms in Gkeyll to investigate edge and SOL turbulence.
Previously, we investigated the use of gyrokinetic continuum algorithms in a 1D1V SOL with logical sheath boundary conditions \citep{Parker1993}
(using one and later two velocity dimensions) with encouraging results \citep{Shi2015}.

We are developing Gkeyll with a number of algorithmic choices to try to better
handle some of the numerical challenges of the edge region.
In the course of developing the code, we ran into and fixed problems
related to some of the above challenges.
For gyrokinetics, Gkeyll uses a full-$f$ formulation with no assumption of
small amplitude fluctuations.
Energy conservation is more difficult to achieve for kinetic Vlasov-type equations than for fluid equations.
The gyrokinetic model in Gkeyll is implemented using a discontinuous Galerkin (DG) algorithm that conserves (in the continuous
time or implicit limit) not only particles but also energy for Hamiltonian terms \citep{Liu2000},
even if limiters \citep{Dumbser2008,LeVeque2002,Durran2010} are applied to the fluxes at cell boundaries.
Limiters on boundary fluxes can only ensure positivity of cell averages \citep{Zhang2010} and we
implement other limiters with correction steps to preserve positivity everywhere within a cell.
There are other possible ways to further improve the DG algorithm in the code
(including exponential basis functions and sparse-grid-quadrature methods)
that can be considered in the future.

In this paper, we detail our numerical approach and present results from gyrokinetic continuum
simulations of electrostatic plasma turbulence in the Large Plasma Device (LAPD) at UCLA \citep{Gekelman1991,Gekelman2016}
using drift-kinetic electrons (with a reduced mass ratio) and gyrokinetic ions.
The LAPD is a linear device that creates a plasma in a straight, open-field-line configuration.
Despite its relatively low plasma temperature, the LAPD contains the basic elements
of a SOL in a simplified (no X-point geometry, straight magnetic field lines, etc.), well-diagnosed
setting, making this device a useful benchmark of edge gyrokinetic algorithms.
The LAPD plasma's relatively high collisionality also facilitates comparisons with Braginskii fluid codes,
where good agreement between the two approaches is expected.

Our work is a gyrokinetic extension of fluid simulations of \citet{Rogers2010} and \citet{Popovich2010a},
and in particular we follow much of the same simulation set up as in \citet{Rogers2010}.
To our knowledge, these are the first 5D gyrokinetic continuum simulations on open-field-lines
including interactions with sheath losses, and in particular are the first 5D gyrokinetic simulations
of a basic laboratory plasma experiment including a sheath model. 
We have also performed simulations of simple magnetized tori, in which the magnetic field lines are helical and the 
magnetic curvature drift is present,
but we defer discussion of those results to a future publication.

\section{Model \label{sec:model}}
We solve the full-$f$ gyrokinetic equation written in the conservative form in the long-wavelength, zero-Larmor-radius limit \citep{Brizard2007,Sugama2000,Idomura2009}
\begin{equation}
  \frac{\partial \mathcal{J}_s f_s}{\partial t} + \nabla \cdot (\mathcal{J}_s \dot{\boldsymbol{R}} f_s) +
  \frac{\partial}{\partial v_\parallel}(\mathcal{J}_s \dot{v}_\parallel f_s) = \mathcal{J}_s C[f_s] + \mathcal{J}_s S_s, \label{eq:gke}
\end{equation}
where $f_s = f_s(\boldsymbol{x}, v_\parallel, \mu, t)$ is the gyrocentre distribution function for species $s$,
$\mathcal{J}_s = m_s^2 B_\parallel^*$ is the Jacobian of the gyrocentre coordinates,
$B_\parallel^* = \boldsymbol{b} \cdot \boldsymbol{B^*}$, $\boldsymbol{B^*} = \boldsymbol{B} + (B v_\parallel/\Omega_i) \nabla \times \boldsymbol{b}$,
$C[f_s]$ represents the effects of collisions, $\Omega_i = eB/m_i$, and $S_s = S_s(\boldsymbol{x}, v_\parallel,\mu, t)$ represents plasma sources (e.g.
neutral ionization or core plasma outflow).
The phase-space advection velocities are defined as
$\dot{\boldsymbol{R}} = \{\boldsymbol{R}, H\}$ and $\dot{v_\parallel} = \{v_\parallel, H\}$,
where the gyrokinetic Poisson bracket operator is
\begin{equation}
\{F,G\} = \frac{\boldsymbol{B^*}}{m_s B_\parallel^*} \cdot \left( \nabla F \frac{\partial G}{\partial v_\parallel} - \frac{\partial F}{\partial v_\parallel} \nabla G \right)
- \frac{1}{q_s B_\parallel^*} \boldsymbol{b} \cdot \nabla F \times \nabla G.
\end{equation}

The gyrocentre Hamiltonian is
\begin{equation}
H_s = \frac{1}{2} m_s v_\parallel^2 + \mu B + q_s \langle \phi \rangle_\alpha, \label{eq:hamiltonian}
\end{equation}
where $\langle \phi \rangle_\alpha$ is the gyro-averaged potential ($\langle \phi \rangle_\alpha = \phi$ in the zero-Larmor-radius limit).

The potential is solved for using the gyrokinetic Poisson equation with a linear ion polarization density
\begin{equation}
-\nabla_\perp \cdot \left( \frac{n_{i0}^g e^2 \rho_{\mathrm{s}0}^2}{T_{e0} } \nabla_\perp \phi \right) =  \sigma_g = e \left[ n_i^g(\boldsymbol{x}) - n_e(\boldsymbol{x}) \right], \label{eq:gkp}
\end{equation}
where $\rho_{\mathrm{s}0} = c_{\mathrm{s}0} / \Omega_i$, $c_{\mathrm{s}0} = \sqrt{T_{e0}/m_i}$, and $n_{i0}^g$ is the background
ion guiding centre density that we will take to be a constant in space and in time.
The replacement of $n_{i}^g(\boldsymbol{x})$ by $n_{i0}^g$ on the left-hand side of (\ref{eq:gkp})
is analogous to the Boussinesq approximation employed in some Braginskii fluid codes.
Future work could generalize this $n_{i0}^g$ to the full $n_i^g(\boldsymbol{x})$ and retain a second-order contribution
to the Hamiltonian (\ref{eq:hamiltonian}) to preserve energy conservation \citep{Krommes2012,Krommes2013,Scott2010}.
Since this paper focuses on simulations of a linear device, the calculations are done in a Cartesian geometry with $x$ and $y$ being
used as coordinates perpendicular to the magnetic field, which lies solely in the $z$ direction.
Therefore, $\nabla_\perp = \boldsymbol{\hat{x}} \partial_x  + \boldsymbol{\hat{y}} \partial_y$.
Note that (\ref{eq:gkp}) is statement of quasineutrality, where the right-hand side is the
guiding-centre component of the charge density, and the left-hand side is the negative of the
ion-polarization charge density, $-\sigma_{\rm pol}$ (due to the plasma response to a cross-field electric field),
so this equation is equivalent to $0 = \sigma = \sigma_g + \sigma_{\rm pol}$.

Electron-electron and ion-ion collisions are implemented using a Lenard-Bernstein model collision operator \citep{Lenard1958}
\begin{eqnarray}
C_{ss}[f_s] &=& \nu_{ss} \frac{\partial}{\partial \boldsymbol{v}} \cdot \left[ (\boldsymbol{v} - \boldsymbol{u}_s) f_s + v_{t,ss}^2 \frac{\partial f_s}{\partial \boldsymbol{v}}\right] \nonumber\\
&=& \nu_{ss} \frac{\partial}{\partial v_\parallel} \left[ (v_\parallel - u_{\parallel,s}) f_s + v_{t,ss}^2 \frac{\partial f_s}{\partial v_\parallel}\right]
+ \nu_{ss} \frac{\partial}{\partial \mu}\left[2\mu f_s + 2 \frac{m_s v_{t,ss}^2}{B} \mu \frac{\partial f_s}{\partial \mu} \right],
\end{eqnarray}
where standard expressions are used for collision frequency $\nu_{ss}$ \citep[p. 37]{Huba2013}, $n_s v_{t,ss}^2 = \int \mathrm{d}^3 v \, m_s \left( \boldsymbol{v} - \boldsymbol{u_s}\right)^2 f_s /3$,
and $n_s u_{\parallel,s} = \int \mathrm{d}^3v \, v_\parallel f_s$.
This collision operator relaxes to a local Maxwellian, contains pitch-angle scattering, and conserves number, momentum, and energy.
Note that the collision frequency is independent of velocity; the $v^{-3}$ dependence of the collision frequency expected for Coulomb collisions
is neglected.
Future work will implement a more sophisticated collision operator,
but this model operator represents many of the key features of the full operator,
including velocity-space diffusion that preferentially damps small velocity-space scales.

Electron-ion collisions are implemented in a similar manner as same-species collisions using the collision operator
\begin{eqnarray}
C_{ei}[f_e] = \nu_{ei} \frac{\partial}{\partial v_\parallel} \left[ (v_\parallel - u_{\parallel,i}) f_e + v_{t,ei}^2 \frac{\partial f_e}{\partial v_\parallel}\right]
+ \nu_{ei} \frac{\partial}{\partial \mu} \left[2\mu f_e + 2 \frac{m_e v_{t,ei}^2}{B} \mu \frac{\partial f_e}{\partial \mu} \right],
\end{eqnarray}
where $\nu_{ei} = \nu_{ee}/1.96$ was used in the simulations and $n_e v_{t,ei}^2 = \int \mathrm{d}^3 v \, m_e \left( \boldsymbol{v} - \boldsymbol{u_i}\right)^2 f_e /3$.
The purpose of this operator is to model the collisional drag and pitch angle scattering of electrons on ions.
The $C_{ie}$ operator is much smaller and is therefore neglected in this work, which results in a small $\mathcal{O}(m_e/m_i)$
violation of momentum conservation.

\subsection{Numerical algorithms}
An energy-conserving (in the continuous-time limit) discontinuous Galerkin algorithm \citep{Liu2000}
is used to discretize the equations in space.
Although \citet{Liu2000} presented their algorithm for the two-dimensional incompressible Euler and Navier-Stokes equations,
we recognized the general applicability of their algorithm for Hamiltonian systems.
Upwind interface fluxes are used in (\ref{eq:gke}) (interface flux terms appear after integrating by parts the product of
(\ref{eq:gke}) and a test function).
This algorithm requires that the Hamiltonian be represented on a continuous subset of the basis set used to represent
the distribution function.
Therefore, the distribution function is represented using discontinuous ($C^{-1}$) polynomials, while the electrostatic
potential is represented using continuous ($C^0$) polynomials (equivalent to continuous finite elements).
Although a single non-local solve is required for the 3D potential, the 5D gyrokinetic equation itself
can be solved in a highly local manner.
Second-order derivatives, which are present in the collision operator,
are calculated using the recovery-based discontinuous Galerkin method \citep{VanLeer2005},
which has the desirable property of producing symmetric solutions.
Time-stepping is performed using an explicit third-order strong-stability-preserving Runge-Kutta algorithm \citep{Gottlieb2001}.

For simplicity, we use nodal, linear basis functions to approximate the solution in each element.
This leads to $32$ degrees of freedom per cell in the 5D phase-space mesh ($8$ degrees of freedom in the 3D configuration-space mesh).
With the $32$ degrees of freedom specified in a cell, the value of $f$ can be computed anywhere within the cell without
additional approximation.
Integration is performed using Legendre-Gauss quadrature, ensuring that the number of quadrature points used is sufficient to
evaluate every required integral exactly.
We use rectangular meshes with uniform cell spacing, but we note that the DG algorithms used are also applicable to
non-uniform and non-rectangular meshes.

\subsubsection{Positivity of the distribution function}
We found it necessary to adjust the distribution function of each species at every time step so that $f_s \ge 0$ at
every node to avoid stability issues.
After much investigation, the main source of negativity in the distribution function appears to be the collision operator at locations 
where the perpendicular temperature of the distribution function is
close to the lowest perpendicular temperature that can be represented on the grid.

If one considers a velocity-space grid made of uniform cells with widths $\Delta v_\parallel$ and $\Delta \mu$ in the
parallel and perpendicular coordinates, the minimum temperatures for a realizable distribution
are computed by assuming that the distribution function
is non-zero at the node located at $(v_\parallel = 0, \mu = 0)$ and 0 at all other nodes.
Using piecewise-linear basis functions,
\begin{eqnarray}
T_{\parallel,\mathrm{min}} &=& \frac{m}{6} \left(\Delta v_\parallel \right)^2 \\
T_{\perp,\mathrm{min}} &=& \frac{B}{3} \Delta \mu \\
T_{\mathrm{min}} &=& \frac{1}{3} \left(T_{\parallel,\mathrm{min}} + 2 T_{\perp,\mathrm{min}} \right)
\end{eqnarray}
Typical values of $\Delta v_\parallel$ and $\Delta \mu$ for a uniformly spaced grid that contains a few $v_{t} = \sqrt{T/m}$
can result in $T_{\parallel,\mathrm{min}} < T_{\perp,\mathrm{min}}$.
A situation can occur in which the collision operator will try to relax the $T_\perp$ at a location to a value below $T_{\perp,\mathrm{min}}$,
resulting in negative-valued regions appearing in the distribution function.

This positivity issue can altogether be avoided by choosing a velocity-space grid that has
$T_{\parallel,\mathrm{min}} = T_{\perp,\mathrm{min}} = T_{\mathrm{min}}$, either by increasing the resolution in $\mu$
relative to the resolution in $v_\parallel$, using a non-uniformly spaced grid in $\mu$,
using non-polynomial basis functions \citep{Yuan2006} that guarantee the positivity of the distribution function,
or using $\sqrt{\mu}$ as a coordinate instead of $\mu$.
We have also developed a DG algorithm that uses exponential basis functions
and conserves energy, and have carried out tests of it in 1D, but leave full implementation of it to future work.
For now, we use a simpler correction procedure described in this section,
which has a philosophy similar to the correction operator used by \citet{Taitano2015}.
The magnitude of the correction operator scales with the truncation error of the method,
and so it vanishes as the grid is refined and does not affect the order of accuracy of the
algorithm while making the simulation more robust on coarse grids by preserving key conservation laws.

For use in these initial simulations, we developed a relatively simple positivity-adjustment procedure
to eliminate the negative-valued nodes of the distribution functions,
while keeping the number density and thermal energy unchanged.
First, the number density, parallel energy, perpendicular energy, and parallel momentum for each species are computed.
Next, all negative-valued nodes of the distribution functions are set to zero, resulting in changes
to the thermal energy and density at locations where the distribution functions have been modified.
To compensate for the increased density, the distribution function is scaled uniformly in velocity space at each configuration space node
to restore the original density.

The remaining task is to modify the distribution function so that no additional energy is added through the positivity-adjustment procedure.
To remove parallel thermal energy $ \int \mathrm{d}^3 v\, \frac{1}{2} m_s v_\parallel^2 f_s $ added through the positivity-adjustment
procedure, we use a numerical drag term of the form
\begin{equation}
\frac{\partial f}{\partial t} = \frac{\partial}{\partial v_\parallel} \left[\alpha_{\mathrm{corr},v_\parallel} \left(v_\parallel - u_\parallel \right) f \right],
\end{equation}
where $\alpha_{\mathrm{corr},v_\parallel}$ is a small numerical correction drag rate that is chosen
each time step to remove the extra parallel energy added.
To guarantee that the numerical drag term will not cause any nodes to go negative,
this operator is implemented in a finite-volume sense, adjusting the mean values:
\begin{equation}
\frac{\bar{f}_j^{n+1} - \bar{f}_j^{n}}{\Delta t} = \frac{\alpha_{\mathrm{corr},v_\parallel}}{\Delta v_\parallel} \left( (v_\parallel-u_\parallel)_{j+1/2} \hat{f}^n_{j+1/2} - (v_\parallel-u_\parallel)_{j-1/2} \hat{f}^n_{j-1/2} \right),
\end{equation}
where the interface flux $\hat{f}^n_{j+1/2} = g(\bar{f}^n_{j}, \bar{f}^n_{j+1})$ is chosen in an upwind sense according to the sign of $v_\parallel-u_\parallel$
and $\bar{f}_j$ denotes the cell-averaged value of $f_j$.
To ensure that the parallel drag term does not modify the perpendicular energy $ \int \mathrm{d}^3 v\, \frac{1}{2} m_s v_\perp^2 f_s $,
this operator is applied at fixed $(\boldsymbol{x},\mu)$.
In our tests, we found that $\alpha_{\mathrm{corr},v_\parallel}$ can not be generally chosen to restore the parallel thermal energy
at every position-space node, since there is a limit on how large $\alpha_{\mathrm{corr},v_\parallel}$
can be while keeping $\bar{f}_j \geq 0$ in every cell.
Instead, we choose $\alpha_{\mathrm{corr},v_\parallel}$ to restore the cell-averaged parallel energy
\begin{equation}
\bar{W}_{\parallel,j} = \int_{x_j - \Delta x/2}^{x_j + \Delta x/2} \mathrm{d}x
\int_{y_j - \Delta y/2}^{y_j + \Delta y/2} \mathrm{d}y
\int_{z_j - \Delta z/2}^{z_j + \Delta z/2} \mathrm{d}z \int \mathrm{d}^3 v\, \frac{1}{2} m_s v_\parallel^2 f_s,
\end{equation}
which results in some position-space diffusion of energy.

We employ a similar procedure to remove the unphysical perpendicular energy added through positivity:
\begin{equation}
\frac{\partial f}{\partial t} = \frac{\partial}{\partial \mu} \left( 2 \alpha_{\mathrm{corr},\mu} \mu f \right).
\end{equation}
Here, the factor $\alpha_{\mathrm{corr},\mu}$ is chosen to restore the cell-averaged perpendicular energy.
Similarly, this operation modifies the perpendicular energy without changing the parallel energy.
Generally speaking, all of the parallel energy added through positivity can usually be removed
through the numerical drag operator while a small amount ($<10\%$) of perpendicular energy added
through positivity remains even after applying the numerical drag operator, a consequence from the choice of a
uniformly spaced grid in $\mu$ (energy is typically added in the distribution function tails, so a uniformly spaced energy grid will
be more constrained than a quadratically spaced energy grid in removing positivity-added energy using a numerical drag operator).
We observe that the cells in which some extra energy is added through the positivity-adjustment procedure
are located on the boundaries in the parallel direction, so we do not expect the extra energy added to have
a significant impact on the quantities of interest in the simulation.

\section{Sheath boundary conditions \label{sec:sheath}}
Debye sheaths form at the plasma-material interface, such as where open magnetic field lines intersect a divertor or limiter.
The sheath width is of order the Debye length and forms on a time scale of order the plasma period,
which are both very disparate scales compared to the turbulence scales of interest in gyrokinetics,
so it is natural and desirable to treat the sheath through model boundary conditions
to avoid the need to directly resolve it.
For example, the Debye length in LAPD is ($\sim 10^{-5}$ m), which is very small compared
to the gyroradius ($\sim 10^{-2}$ m) and even smaller compared to the parallel scales of
the turbulence ($\sim 10$ m).
The plasma frequency is $\sim 10^{11}$ s$^{-1}$, which is much larger compared to the ion gyrofrequency
($\sim 10^{6}$ s$^{-1}$), and even larger than the turbulence frequencies of interest
($ \omega_* \sim 10^{4}$ s$^{-1}$ at $k_\theta \rho_{\mathrm{s}0} \sim 0.3$).
Furthermore, the quasineutrality and low-frequency assumptions of gyrokinetics break
down in the sheath, so gyrokinetic models cannot directly handle sheaths.


We use (\ref{eq:gkp}) to solve for the potential $\phi(x,y,z)$ everywhere in the simulation domain.
The sheath potential $\phi_{sh}(x,y)$ on each boundary in $z$ (where the field lines intersect the wall)
is obtained by simply evaluating $\phi$ on that boundary, so at the lower boundary, $\phi_{\rm sh}(x,y) =  \phi(x,y,-L_z/2)$.
The wall is taken to be just outside the simulation domain and the wall potential $\phi_w$ is 0 for a grounded wall.
Outgoing particles with $\frac{1}{2} m_s v_\parallel^2 < -q_s \left( \phi_{sh}-\phi_w \right)$ are reflected (e.g. when $\phi_{sh}$ is positive,
some electrons will be reflected), while the rest of the outgoing particles leave the simulation domain.
This procedure is analogous to how some fluid codes determine $\phi$ everywhere (including the sheath potential) from the fluid vorticity equation
and then use the sheath potential to set the boundary condition on the parallel electron velocity (sometimes called a conducting-wall boundary condition)
\citep{Xu1998,Rogers2010,Friedman2013}.

Note that our present sheath model for electrons is different than the logical sheath model \citep{Parker1993},
which determines the sheath potential each time step by requiring that the electron
flux match the ion flux at each point on the wall so there is no current to the wall
(this might be considered a model for an insulating wall).
In the present conducting wall approach, the sheath potential is determined by other
effects (the gyrokinetic Poisson equation or the related fluid vorticity equation),
and then used to determine what fraction of electrons are reflected and thus the
resulting currents to the wall.
If one starts with an initial condition where $\sigma_g=0$ in (\ref{eq:gkp}) so $\phi=0$,
then electrons will rapidly leave the plasma, causing the guiding centre charge $\sigma_g$
to rise to be positive, and thus the sheath potential will quickly rise to reflect most of the electrons,
and bring the sheath currents down to a much smaller level,
while allowing the sheath currents to self-consistently fluctuate in interactions with the turbulence.
Currents are allowed to flow in and out of the wall, with current paths closing through the wall.

In the code, this reflection procedure is applied at each node on the upper and lower surfaces in $z$
at the end of the simulation domain $z=\pm L_z/2$ (adjacent to the end plates), where
the reflected distribution function $f_R(\boldsymbol{x},v_\parallel,\mu)$ is set in ghost cells.
Let's consider a case in which $\phi_{sh}-\phi_w$ is positive on a node in the upper $z$ boundary, so low-energy outgoing electrons with
$0 < v_\parallel < v_{\mathrm{cut},e} = \sqrt{2 e \left(\phi_{sh}-\phi_w\right)/m_e}$ are reflected
with velocity $-v_\parallel$ and all outgoing ions leave the simulation domain.

Since the distribution function is discretized on a phase-space grid, each cell is associated with a range of parallel velocities
$v_{c,j} - \Delta v_\parallel/2 < v_\parallel < v_{c,j} + \Delta v_\parallel/2$, where $v_{c,j}$ is the $v_\parallel$ coordinate of 
the centre of cell $j$ and $\Delta v_\parallel$ is the width of cell $j$ in the $v_\parallel$ direction.
For cells whose parallel velocity extents do not bound $v_{\mathrm{cut},e}$, the reflection procedure is straightforward:
find the corresponding ghost cell $j'$ with $v_{c,j'} = -v_{c,j}$ and copy the solution after reflection about the $v_\parallel$ axis.

In the cells whose parallel velocity extents bound $v_{\mathrm{cut},e}$, the distribution function copied into the corresponding ghost cell
needs to be both reflected about the $v_\parallel$ axis and scaled by a factor so that the net outward flux has the correct value
based on the reflection of outgoing particles with $v_\parallel < v_{\mathrm{cut},e}$.
Due to the numerical representation of the distribution function, which is a local polynomial expansion in each configuration space cell,
it is not possible to represent a reflected distribution function that is zero for all $v_\parallel < - v_{\mathrm{cut},e}$ unless $v_{\mathrm{cut},e}$ happens
to lie on the boundary between two cells.
Therefore, the reflected distribution function in the cutoff cell is scaled by the fraction
\begin{equation}
c = \frac{\int_{v_j - \Delta v/2}^{v_{\mathrm{cut},e}} \mathrm{d}v_\parallel \int_0^{\mu_\mathrm{max}} \mathrm{d}\mu \, v_\parallel f_e}
{\int_{v_j - \Delta v_\parallel/2}^{v_j + \Delta v_\parallel/2} \mathrm{d}v_\parallel \int_0^{\mu_\mathrm{max}} \mathrm{d}\mu \, v_\parallel f_e},
\end{equation}
although this is just one of many choices in modifying the reflected distribution function so that the net outward flux has the correct value.

So far, we have only described the boundary condition for the electrons.
The boundary condition we use for ions is the same as the one used in the logical sheath model \citep{Parker1993}
(a variant of which is used in the XGC gyrokinetic PIC code \citep{Churchill2016}):
the ions just pass out freely at whatever velocity they have been accelerated to
by the potential drop from the upstream source region to the sheath entrance.
(This is for a normal positive sheath. In the unusual situation that the sheath potential
were to go negative, then some ions would be reflected.)
The only boundary condition that the sheath model imposes on the ions is that there are no incoming ions,
i.e., at the incoming lower sheath boundary we have the boundary condition that
$f_i(x,y,z=-L_z/2,v_\parallel, \mu) = 0$ for all $v_{\parallel} \ge 0$.
While this leads to a well-posed set of boundary conditions, and appears to work well and give
physically reasonable results for the simulations carried out in this paper,
it might need improvements in some parameter regimes.
These issues will be considered in future work.

\subsection{Future considerations for sheath models}
Sheaths have long been studied in plasma physics, including kinetic effects
and angled magnetic fields, and there is a vast literature on them.
The standard treatments look at steady-state results in 1D, in which the potential
is determined by solving the Poisson equation along a field line
(for the case here in which the magnetic field is perpendicular to the surface),
but for gyrokinetic turbulence, we need to consider time-varying
fluctuations in which the sheath region needs to couple to an upstream
gyrokinetic region where the potential is determined in 2D planes
perpendicular to the magnetic field by solving the gyrokinetic quasineutrality
equation (\ref{eq:gkp}).
The details of how this matching or coupling is carried out may depend on the particular
numerical algorithm used and how it represents electric fields near a boundary.

There are a range of possible sheath models of different levels of
complexity and accuracy that could be considered in future work.
The present model does not guarantee that the Bohm sheath criterion is met,
which requires that the ion outflow velocity exceed the sound speed, $u_{\parallel i} \ge c_\mathrm{s}$,
for a steady-state sheath and in the sheath-entrance region.
However, the present simulations start at a low density and ramp up
the density to an approximate steady state over a period of a few sound transit times,
and during this phase, the pressure and potential drop from the central
source region to the edges is large enough to accelerate ions to supersonic velocities.
(As we will see in figure \ref{fig:lapd2dxz}, the potential drop from the centre of the
simulation to the edge in $z$ is larger than the electron temperature near the edge.)

There could be other cases where the acceleration of ions in the upstream region
is not strong enough to enforce the Bohm sheath criterion for a steady-state result.
In such a case, some kind of rarefaction fan may propagate from near the sheath,
accelerating ions back up to a sonic level.
This situation is very similar to the Riemann problem for the expansion of a gas into a vacuum \citep{Munz1994}
or into a perfectly absorbing surface, which leads to a rarefaction wave that always maintains
$u_{|| i} \ge c_\mathrm{s}$ at the boundary (but also modifies the density and temperature at the outflow
boundary because of the rarefaction in the expanding flow).
A Riemann solver has been implemented in the two-fluid version of Gkeyll for 1D simulations that resolve the sheath 
\citep{Cagas2016}, and the results were compared with a fully kinetic solver.
Exact and approximate Riemann solvers are often used in computational fluid dynamics to
determine upwind fluxes at an interface \citep{LeVeque2002,Durran2010}.
It could be useful to work out a kinetic analogue of this process, or a kinetic model based
on the approximate fluid result, but those are beyond the scope of this paper.

There is ongoing research to develop improved sheath models for fluid codes.
In some past fluid simulations of LAPD, the parallel ion dynamics was neglected
and modelled by sink terms to maintain a desired steady-state on average \citep{Popovich2010b,Friedman2012,Friedman2013}.
Rogers and Ricci included parallel ion dynamics in their fluid simulations \citep{Rogers2010,Ricci2010}
and imposed the boundary condition $u_{\parallel i} = c_\mathrm{s}$, thus avoiding the
problem of $u_{\parallel i} < c_\mathrm{s}$.
This could be generalized in the future to allow
$u_{\parallel i} > c_\mathrm{s}$ at the boundary to handle cases where turbulent fluctuations
or other effects give more upstream acceleration \citep{Togo2016,Dudson2016}.
\citet{Loizu2012} carried out a kinetic study to develop improved sheath model boundary conditions
for fluid codes that include various effects (including the magnetic pre-sheath \citep{Chodura1982} in an oblique magnetic
field and the breakdown of the ion drift approximation) that have been incorporated
into later versions of the GBS code.

\section{Simulations of LAPD \label{sec:results} }
We selected the parameters for our simulations of a LAPD-like plasma based on those used
by \citet{Rogers2010} in a previous Braginskii-fluid-based study, with some modifications for use in a kinetic model:
$T_{e0} = 6$ eV, $T_{i0} = 1$ eV, $m_i = 3.973 m_p$ ($m_p$ is the proton mass), $B = 0.0398$ T, and
$n_0 = 2 \times 10^{18}$ m$^{-3}$.
As done by \citet{Rogers2010}, we have also used a reduced mass ratio of $m_e/m_i = 1/400$, which allows for larger time steps to be taken,
but weakens the adiabatic electron response.
We have also reduced the electron-electron collision frequency by a factor of 10 for these simulations,
which increases the minimum stable explicit-time-step size while keeping the collisional mean free path small compared to
the parallel length of the simulation box ($v_{t,e}/\nu_{ee} \sim 10^{-2}$ m for typical parameters).
In the future, we plan to implement an implicit or super-time-stepping algorithm
for the collision operator to be able to take much larger time steps with the physical collision frequency.
The rectangular simulation box (an approximation to the cylindrical LAPD plasma) has perpendicular lengths
$L_\perp = L_x = L_y = 100 \rho_{\mathrm{s}0}$ and parallel length $L_z = 18$ m, where
$\rho_{\mathrm{s}0} = c_{\mathrm{s}0}/\Omega_i$ and $c_{\mathrm{s}0} = \sqrt{T_{e0}/m_i}$.
The grid parameters are summarized in table \ref{tab:grid}, with 32 degrees of freedom stored in each cell.
With these parameters, $T_{e,\mathrm{min}} = 0.9067$ eV, $T_{\parallel e,\mathrm{min}} = 0.32$ eV,
and $T_{\perp e,\mathrm{min}} = 1.2$ eV.
For time-stepping, the Courant number is set to 0.1.

\begin{table}
\begin{center}
\begin{tabular}{cccc}
Coordinate & Number of Cells & Minimum & Maximum \\
$x$ & 36 & $-50 \rho_{\mathrm{s}0}$ & $50 \rho_{\mathrm{s}0}$ \\
$y$ & 36 & $-50 \rho_{\mathrm{s}0}$ & $50 \rho_{\mathrm{s}0}$ \\
$z$ & 10 & $-L_z/2$ & $ L_z/2$ \\
$v_\parallel$ & 10 & $-4 \sqrt{T_{s,\mathrm{grid}}/m_s} $ & $ 4 \sqrt{T_{s,\mathrm{grid}}/m_s} $ \\
$\mu$ & 5 & 0 & $0.75 \frac{m_s v_{\parallel,\mathrm{max}}^2}{2 B_0}$ \\
\end{tabular}
\caption{Parameters for the phase-space grid used in the LAPD simulations. The temperatures
appearing in the velocity-space extents are
$T_{i,\mathrm{grid}} = 1$ eV and $T_{e,\mathrm{grid}} = 3$ eV.
Piecewise-linear basis functions are used, resulting in 32 degrees of freedom per cell}
\label{tab:grid}
\end{center}
\end{table}

Although we expect the quasisteady state of the system to be insensitive to the choice of initial conditions,
we found that it was important to start the simulation with a non-uniform density profile
to avoid exciting large transient potentials that resulted in extremely small restrictions being imposed on the time step.
Because the boundary conditions force $\phi$ to a constant on the side walls, electrons near the domain boundaries in $x$ and $y$ are quickly lost at thermal speeds
from the simulation box.
We believe that this large momentary imbalance in the electron and ion densities is the source of this stability issue.

The initial density profile for both ions and electrons is chosen to be
$n_0 A(r;c_{\mathrm{edge}}=1/20)$, where $A(r;c_{\mathrm{edge}})$ is a function that falls from the peak value
of $1$ at $r=0$ to a constant value $c_{\mathrm{edge}}$ for $r > L_\perp/2$:
\begin{equation}
A(r;c_{\mathrm{edge}}) = \left\{ 
  \begin{array}{ll}
  (1-c_{\mathrm{edge}}) \left( 1 - \frac{r^2}{(L_\perp/2) ^2} \right)^3 + c_{\mathrm{edge}}, & r < L_\perp/2 \\
  c_{\mathrm{edge}},  & \mathrm{else}.
  \end{array} \right.
\end{equation}
The initial electron temperature profile has the form $5.7 A(r;c_{\mathrm{edge}}=1/5)$ eV,
while the initial ion temperature profile is a uniform 1 eV.
Both electrons and ions are initialized as non-drifting Maxwellians, although
future runs could be initialized with a specified non-zero mean velocity as a function of the parallel coordinate
computed from simplified 1D models  \citep{Shi2015} to reach a quasisteady state more quickly.

The electron and ion sources have the form
\begin{equation}
S_s = 
  1.08 \frac{n_0 c_{\mathrm{s}0}}{L_z} \left\{ 0.01 + 0.99 \left[ \frac{1}{2}-\frac{1}{2} \tanh\left( \frac{r-r_s}{L_s} \right) \right] \right \} F_{M,s}(v_\parallel,\mu; T_s) \label{eq:source},
\end{equation}
where $r = \sqrt{x^2 + y^2}$, $r_s = 20 \rho_{\mathrm{s}0} = 0.25$ m, $L_s = 0.5 \rho_{\mathrm{s}0} = 0.625$ cm, and
$F_{M,s}(v_\parallel,\mu; T_s)$ is a normalized non-drifting Maxwellian distribution for species $s$ with temperature $T_s$.
The ion source has a uniform temperature of 1 eV, while the electron source has a temperature profile given by
$6.8 A(r;c_{\mathrm{edge}}=1/2.5)$ eV.
Unlike the sources used by \citet{Rogers2010},
the sources we use model the neutrals as being ionized at zero mean velocity.
In the fluid equations of \citet{Rogers2010}, a zero-velocity plasma source would give rise to an additional
term $-S_n V_{\parallel i} / n$ on the right-hand side of the $\partial_t V_{\parallel i}$ equation,
which is kept in the more general equations of \citet{Wersal2015}.
In our simulations, electrons and ions are also sourced in the $r > r_s$ region at $1/100$th the amplitude of the central source rate
to avoid potential issues arising from zero-density regions.
While there are no primary electrons in the $r > r_s$ region in the actual LAPD device,
\citet{Carter2009} have discussed
the possibility of ionization in this region from rotation-heated bulk electrons.

\begin{figure}
  \centerline{\includegraphics[width=\textwidth]{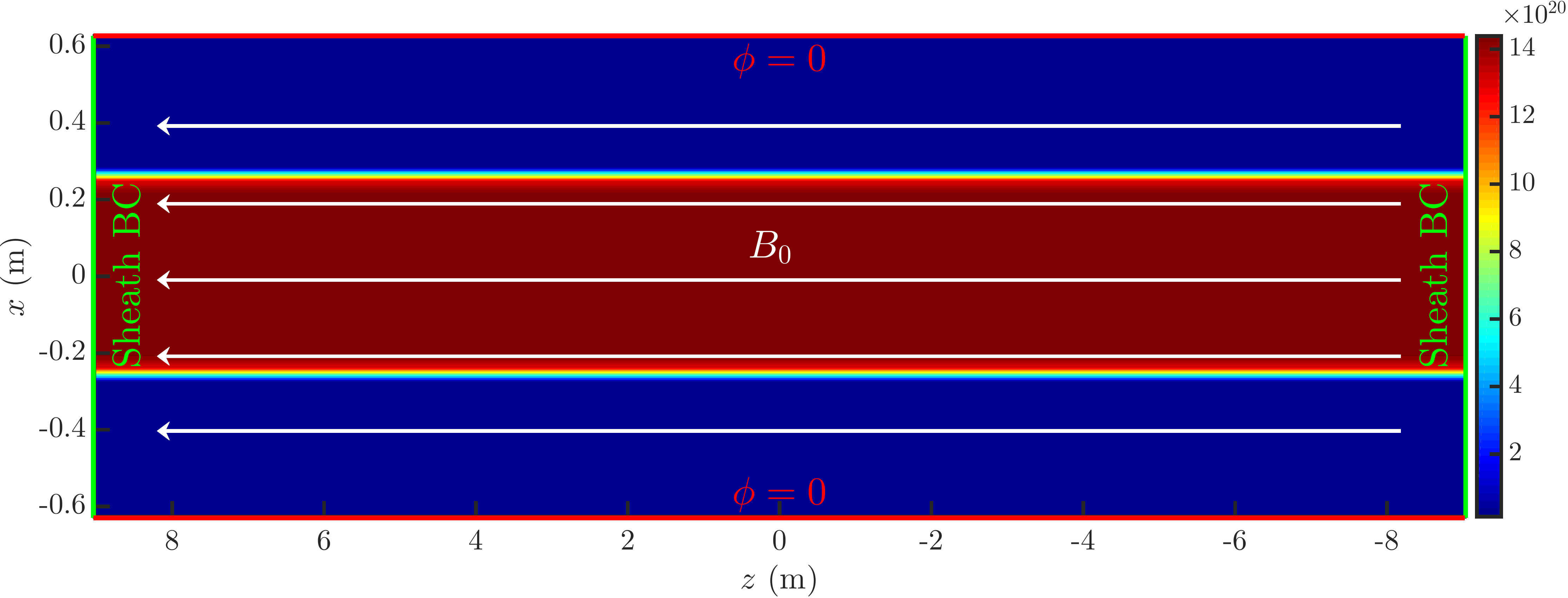}}
  \caption{A plot of the LAPD simulation plasma source (in m$^{-3}$s$^{-1}$) in the $x$-$z$ plane. Annotations indicate the
  direction of the magnetic field, side wall boundary conditions, and sheath boundary condition locations.}
\label{fig:source}
\end{figure}

\subsection{Boundary conditions and energy balance}
Dirichlet boundary conditions $\phi = 0$ are used on the $x$ and $y$ boundaries for the potential solve
(taking the side walls to be grounded to the $\phi_w = 0$ end plates),
while no boundary condition is required in $z$ because (\ref{eq:gkp}) contains no $z$ derivatives.
The distribution function uses zero-flux boundary conditions in $x$, $y$, $v_\parallel$, and $\mu$,
which amounts to zeroing out the interface flux evaluated on a boundary where zero-flux boundary conditions are to be applied.
This ensures that particles are not lost through the domain boundaries in $x$, $y$, $v_\parallel$, and $\mu$.
It should be noted that zero-flux boundary conditions on the $x$ and $y$ boundaries are a result of the choice of a constant $\phi$ on the
side-wall boundaries, so the ExB velocity at these boundaries is parallel to the wall.
Sheath model boundary conditions, discussed in the previous section, are applied on the upper and lower boundaries in the $z$ direction.

To demonstrate how the choice of $\phi = 0$ affects the energy balance in the system,
we define the plasma thermal energy as
\begin{equation}
W_K = \int \mathrm{d}^3x \sum_s \int \mathrm{d}^3 v f_s H_0,
\end{equation}
where $H_0 = \frac{1}{2} m v_\parallel^2 + \mu B$.
Neglecting sources and collisions for simplicity, the kinetic equation in a straight, constant magnetic field can be written as
\begin{equation}
\frac{\partial f_s}{\partial t} + \frac{\partial}{\partial z} \left( v_\parallel f_s \right) + \nabla \cdot \left(\boldsymbol{v}_E f_s \right)
+ \frac{\partial}{\partial v_\parallel} \left( \frac{q_s}{m_s} E_\parallel f_s \right)= 0,\label{eq:gksimple}
\end{equation}
where $E_\parallel = -\boldsymbol{b}\cdot \nabla \langle \phi \rangle$ and $\boldsymbol{v}_E = \mathbf{b} \times \nabla \langle \phi \rangle / B$.

Multiplying (\ref{eq:gksimple}) by $H_0$ and integrating over phase space,
\begin{eqnarray}
\frac{\partial W_K}{\partial t} &=& - \int \mathrm{d}x \, \mathrm{d}y \sum_s \int \mathrm{d}^3 v 
H_0 v_\parallel f_s \Big|_{z_{\mathrm{lower}}}^{z_{\mathrm{upper}}}
+ \int \mathrm{d}^3 x \sum_s \int \mathrm{d}^3 v \, v_\parallel f_s  q_s E_\parallel \nonumber \\
&=& - \int \mathrm{d}x \, \mathrm{d}y \sum_s \int \mathrm{d}^3 v 
H_0 v_\parallel f_s \Big|_{z_{\mathrm{lower}}}^{z_{\mathrm{upper}}} +
\int \mathrm{d}^3 x j_\parallel E_\parallel, \label{eq:w_k}
\end{eqnarray} 
where we have used the fact that the normal component of $\boldsymbol{v}_E$ vanishes on the side walls
(since $\phi$ is a constant on the side walls)
and zero-flux boundary conditions on $f_s$ in $v_\parallel$.
The first term on the right-hand side is the parallel heat flux out to the sheaths and the second term is the parallel acceleration by the electric field,
which mediates the transfer of energy between thermal and field energies in this model (this term appears with the opposite sign in the
equation for the evolution of ExB energy).

To calculate the field energy evolution, we take the time derivative of the gyrokinetic Poisson equation (\ref{eq:gkp}),
\begin{eqnarray}
-\nabla_\perp \cdot \left( \epsilon \nabla_\perp \frac{\partial \phi}{\partial t} \right) &=& \sum_s q_s \int \mathrm{d}^3 v \frac{\partial f_s}{\partial t} \nonumber \\
&=& -\sum_s q_s \int \mathrm{d}^3 v \, \left[ \frac{\partial}{\partial z} \left( v_\parallel f_s \right) + \nabla \cdot \left(\boldsymbol{v}_E f_s \right)
 \right] \nonumber \\
&=& -\frac{\partial j_\parallel}{\partial z} - \nabla \cdot (\boldsymbol{v}_E \sigma_g) \label{eq:gkp_dt},
\end{eqnarray}
where $\epsilon = n_{i0}^g e^2 \rho_{\mathrm{s}0}^2/T_{e0}$.

Next, we multiply (\ref{eq:gkp_dt}) by $\phi$ and integrate over space:
\begin{eqnarray}
-\int \mathrm{d}^3 x \, \phi \nabla_\perp \cdot \left( \epsilon \nabla_\perp \frac{\partial \phi}{\partial t} \right) &=&
-\int \mathrm{d}^3 x \, \phi \left[ \frac{\partial j_\parallel}{\partial z} + \nabla \cdot (\boldsymbol{v}_E \sigma_g) \right] \nonumber \\
- \int \mathrm{d} \boldsymbol{S}_\perp \cdot \phi \epsilon \nabla_\perp \frac{\partial \phi}{\partial t} 
+ \frac{1}{2} \int \mathrm{d}^3 x \, \epsilon \frac{\partial \left( \nabla_\perp \phi \right)^2}{\partial t} &=&
-\int \mathrm{d} x \mathrm{d} y \, \phi j_\parallel \Big|_{z_{\mathrm{lower}}}^{z_{\mathrm{upper}}} + \int \mathrm{d}^3 x \, \frac{\partial \phi}{\partial z} j_\parallel \nonumber\\
&& - \int \mathrm{d} \boldsymbol{S}_\perp \cdot \phi \boldsymbol{v}_E \sigma_g \nonumber \\
&& + \int \mathrm{d}^3 x \, \nabla \phi \cdot \boldsymbol{v}_E \sigma_g \label{eq:gkp_dt_energy}
\end{eqnarray}
The integral involving $\int \mathrm{d} \boldsymbol{S}_\perp$ on the right-hand side is zero because $\boldsymbol{v}_E$ has no normal component on the side walls.
By assuming that $\phi = 0$ on the side walls, the term on the left-hand side involving $\int \mathrm{d} \boldsymbol{S}_\perp \phi$ is also zero and we have
\begin{equation}
W_\phi = \frac{\partial}{\partial t} \left( \frac{1}{2} \int \mathrm{d}^3 x \, \epsilon \left(\nabla_\perp \phi \right)^2 \right) =
-\int \mathrm{d} x \mathrm{d} y \, \phi j_\parallel \Big|_{z_{\mathrm{lower}}}^{z_{\mathrm{upper}}} - \int \mathrm{d}^3 x \, j_\parallel E_\parallel.
\label{eq:w_phi}
\end{equation}
If the wall is biased instead of grounded, as done in a set of experiments by \citet{Carter2009},
one must retain the first term on the left-hand side of (\ref{eq:gkp_dt_energy}) in energy-balance considerations.
The second term on the right-hand side of (\ref{eq:w_phi}) is equal and opposite to the second
term on the right-hand side of (\ref{eq:w_k}), and so cancels when the two equations are added together.
The total energy is the sum of the kinetic energy $W_k$ and the field energy $W_\phi$.
Substituting the definition of $\epsilon$, this field energy can be written as
$W_\phi = \int d^3 x (1/2) n_{i0}^g m_i v_E^2$, indicating that it can be interpreted as
the kinetic energy associated with the ExB motion.
(The $n_{i0}^g$ factor can be generalized to the full density $n_i^g({\boldsymbol{x}},t)$
as described in \S\ref{sec:model}, with an additional contribution to the Hamiltonian.)
The first term on the right-hand side of (\ref{eq:w_phi}) corresponds to work done on particles as they are
accelerated through the sheath. The $\phi$ in this boundary term is the potential at the $z$
boundaries of the simulation domain, where the sheath entrances are.
When $j_{\parallel}=0$ at the sheath entrance, then the energy lost by electrons as they drop
through the sheath is exactly offset by the energy gained by ions as they drop through the sheath.
If more electrons than ions are leaving through the sheath, then the net energy lost in the unresolved sheath
region contributes to an increase in the field energy.

Future studies could also investigate improvements for the side wall boundary conditions.
Identifying the left-hand side of (\ref{eq:gkp_dt}) as $-\partial \sigma_{\mathrm{pol}}/ \partial t = \nabla \cdot \boldsymbol{j}_{\mathrm{pol}}$,
and integrating over all space,
\begin{eqnarray}
\int \mathrm{d}^3 x \, \nabla \cdot \boldsymbol{j}_\mathrm{pol} &=& \int \mathrm{d} \boldsymbol{S} \cdot \boldsymbol{j}_\mathrm{pol} \nonumber \\
&=& \int \mathrm{d} \boldsymbol{S}_\perp \cdot \epsilon \frac{\partial \boldsymbol{E}_\perp}{\partial t},
\end{eqnarray}
so we see that there is an ion polarization current into the side wall when the electric field
pointing into the side wall is increasing in time, which is physically reasonable.
However, if the sign of the electric-field time derivative reverses,
it is not possible to pull ions out of the side wall (where they are trapped by quantum effects, or return as neutrals),
and a boundary layer might form near the side walls.
In fusion devices, it is rare for the magnetic field to be exactly parallel to the wall,
so it it could be appropriate to use a model of the Chodura magnetic pre-sheath \citep{Chodura1982}.
\citet{Geraldini2017} also recently studied a gyrokinetic approach to the magnetic pre-sheath.

The inclusion of charge-neutral source terms and number-conserving collision operators to the above analysis does not result in
additional sources of ExB energy, since they lead to the addition of terms to the right-hand side of (\ref{eq:w_phi}) of the form
\begin{equation}
-\int \mathrm{d}^3x \, \phi \sum_s q_s \int \mathrm{d}^3 v \, S_s (\boldsymbol{x}, \boldsymbol{v}, t) = 0.
\end{equation}

\subsection{Results}
In this section we present results from various quantities derived from our gyrokinetic simulation.
Our goal here is not to argue that our simulations are a faithful model of the LAPD plasma,
but instead to demonstrate the ability to carry out gyrokinetic continuum simulations
of open-field-line plasmas in a numerically stable way and to
demonstrate a reasonable level of qualitative agreement 
by making contact with turbulence measurements from the real LAPD device
and previous Braginskii fluid simulations \citep{Ricci2010,Fisher2015,Friedman2012},
since we have used similar plasma parameters and geometry.
Starting from the initial conditions described in \S\ref{sec:results}, the electron and ion distributions evolve for a few
ion sound transit times ($\tau_s \sim L_z/2/c_s \approx 1.1$ ms using $T_e = 3$ eV) until a quasisteady state is reached, during which the
total number of particles of each species remains approximately constant.

\begin{figure}
  \centerline{\includegraphics[width=\textwidth]{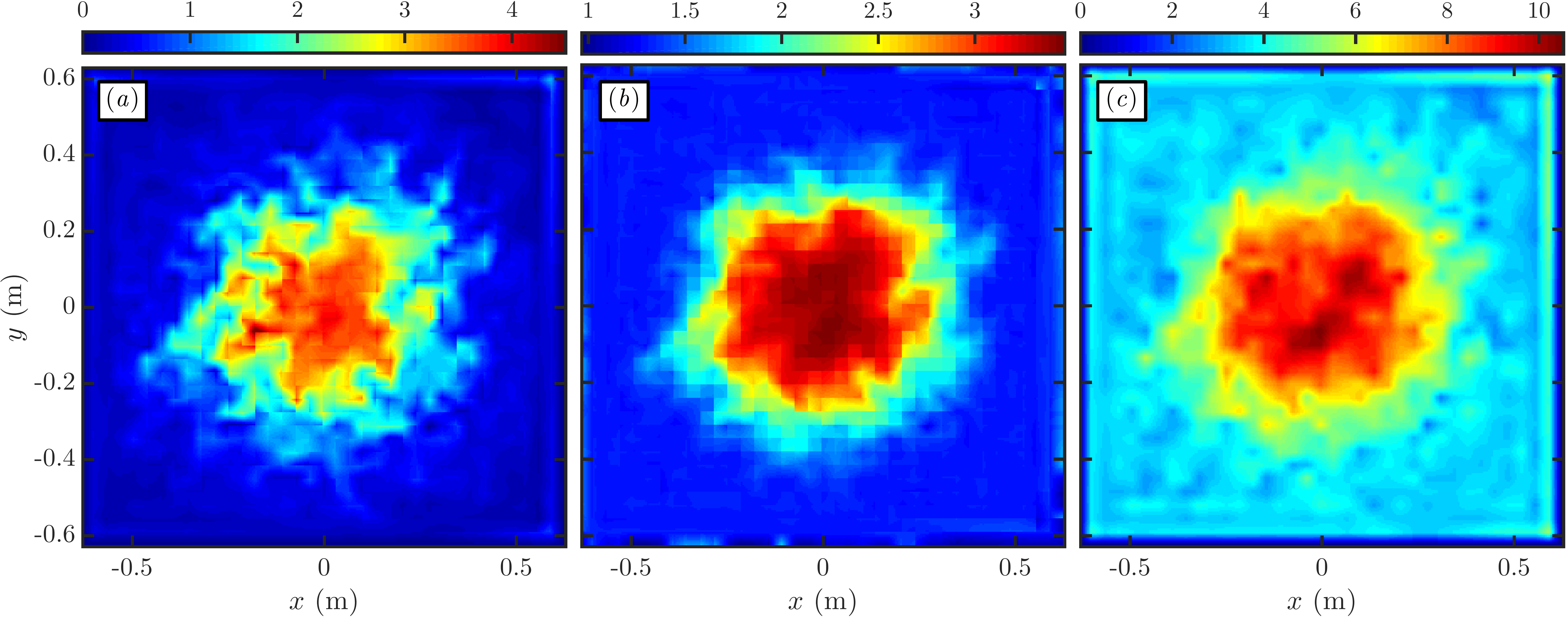}}
  \caption{Snapshots of the (\textit{a}) total electron density (in $10^{18}$ m$^{-3}$), (\textit{b})
  electron temperature (in eV), and (\textit{c}) electrostatic potential (in V) from a 5D gyrokinetic simulation of a turbulent LAPD plasma.
  The plots are made in centre of the box at $z=0$ m.
In this simulation, a continuous source of plasma concentrated inside $r_s = 0.25$ m is transported radially outward by the turbulence
as it flows at near-sonic speeds along the magnetic field lines to the end plates, where losses are mediated by
sheath model boundary conditions.
The plots are made in a plane perpendicular to the magnetic field in the middle of the device after a few ion transit times.}
\label{fig:lapd2dxy}
\end{figure}

\begin{figure}
  \centerline{\includegraphics[width=\textwidth]{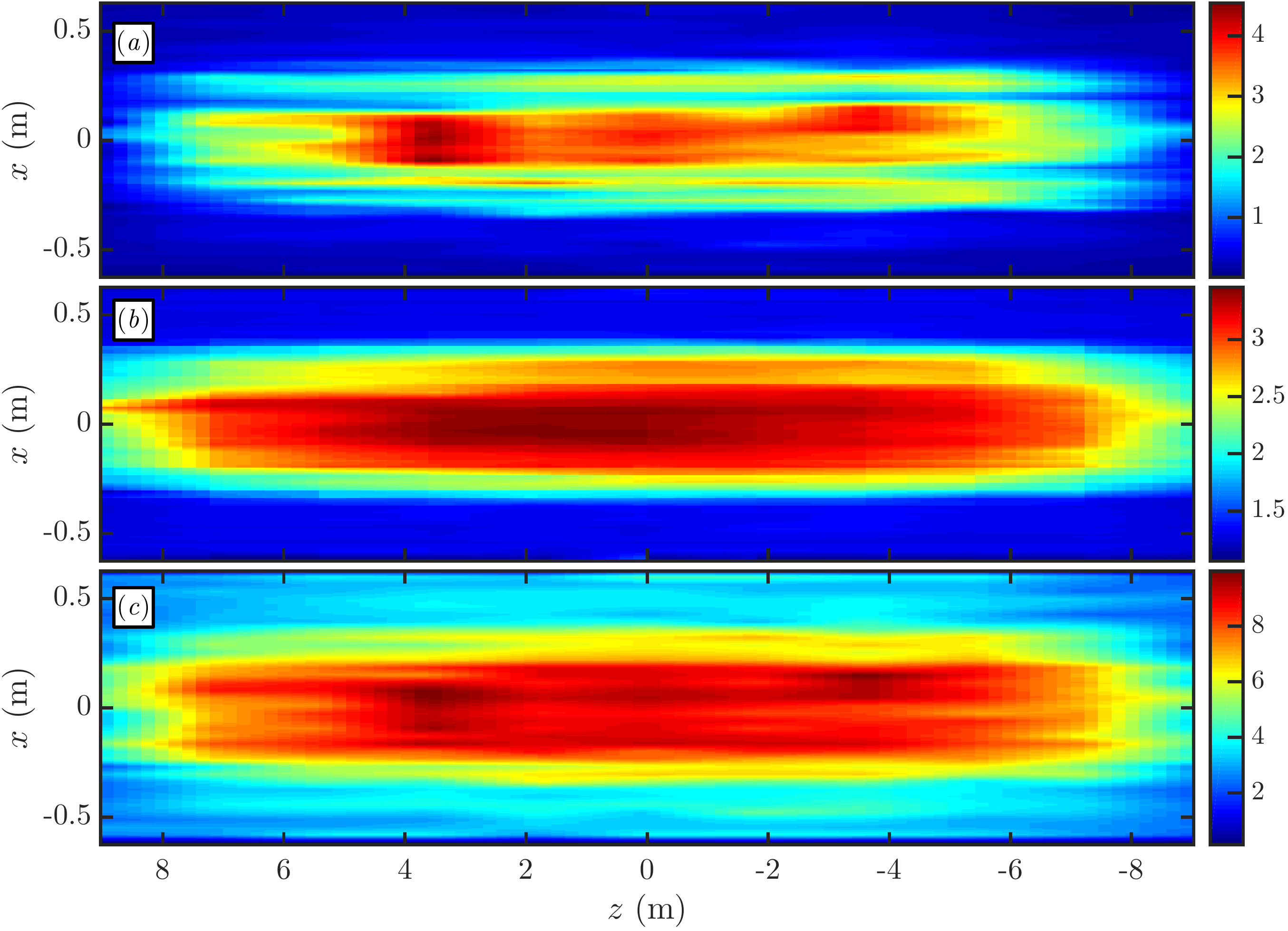}}
  \caption{Snapshots of the (\textit{a}) total electron density (in $10^{18}$ m$^{-3}$), (\textit{b})
  electron temperature (in eV), and (\textit{c}) electrostatic potential (in V) from a 5D gyrokinetic simulation of a turbulent LAPD plasma.
The plots are made in the $(x,z)$ plane at $y=0$ m after a few ion transit times.}
\label{fig:lapd2dxz}
\end{figure}

As seen in LAPD experiments \citep{Schaffner2012,Schaffner2013}, we observe a weak spontaneous rotation in the
ion-diamagnetic-drift direction.
Figure~\ref{fig:lapd2dxy} shows snapshots in the perpendicular plane of the total electron density, electron temperature, and electrostatic potential
after a few ion transit times, which are qualitatively similar to the snapshots presented from
Braginskii fluid simulations of LAPD \citep{Rogers2010,Fisher2015}.
Figure \ref{fig:lapd2dxz} shows the same fields as in figure~\ref{fig:lapd2dxy}, but the plots are made in the $y=0$ plane
to show the parallel structure.

\begin{figure}
  \centerline{\includegraphics[width=\textwidth]{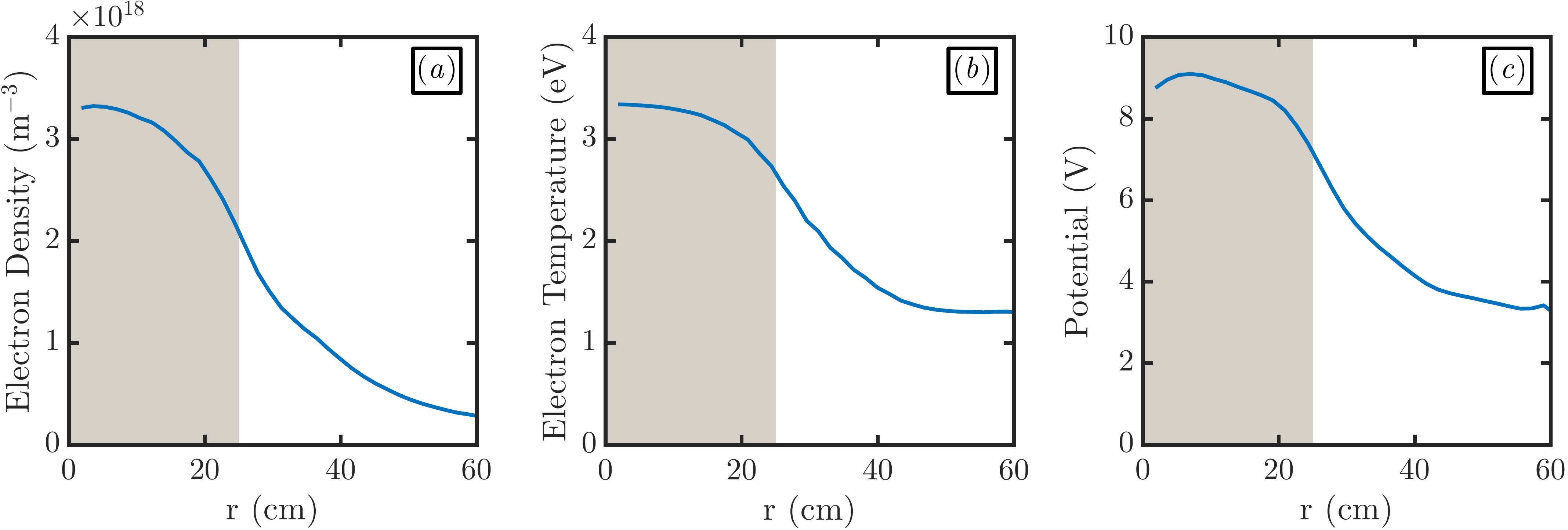}}
  \caption{Plots of the average (\textit{a}) electron density, (\textit{b})
  electron temperature, and potential (\textit{c}) profiles as a function of radius.
  The fields are time-averaged over several ion transit times after the simulation has reached a quasisteady state,
  restricted to $-4$ m $<z< 4$ m,
  evaluated at eight equally spaced points in each cell, and then binned by radius.
  The shaded region in $(a)$ illustrates
  the extent of the strong plasma source.}
\label{fig:lapd1d}
\end{figure}

Figure \ref{fig:lapd1d} shows the time-averaged radial profile of $n_e$, $T_e$, and $\phi$ computed
by averaging the data in the region $-4$ m $<z<4$ m.
We focus on this region since it is similar to the region in which probe measurements are taken in the LAPD,
and there is little parallel variation in this region.
Particle transport in the radial direction is especially evident in figure \ref{fig:lapd1d} from the broadening in the $n_e$ profile.
In figure \ref{fig:lapd1d}, the electron temperature drops off at mid-radii
but is rather flat at large $r$.
To understand this, note that there is a 2.72 eV residual electron source at large $r$ (see (\ref{eq:source})),
and that the observed temperature is close to the limit of the coldest temperature that can be represented on
the grid when collisions dominate and the distribution function is isotropic,
so $T_{e,\mathrm{min}} \sim T_{\perp e,\mathrm{min}} = 1.2$ eV.
Our choice of velocity-space grid is a compromise between resolving low energies and
the need to go up to significantly higher energies than the temperature of
the source (which has a maximum temperature of 6.7 eV) to represent the tail.
This will be improved in future work using a non-uniformly spaced velocity grid or exponential
basis functions, which can represent a range of electron energies much more efficiently.
We do not expect the non-vanishing $T_e$ at large $r$ to affect the results significantly
because both $n_e$ and the $n_e$ fluctuation level are small at large $r$.

\begin{figure}
  \centerline{\includegraphics[width=\textwidth]{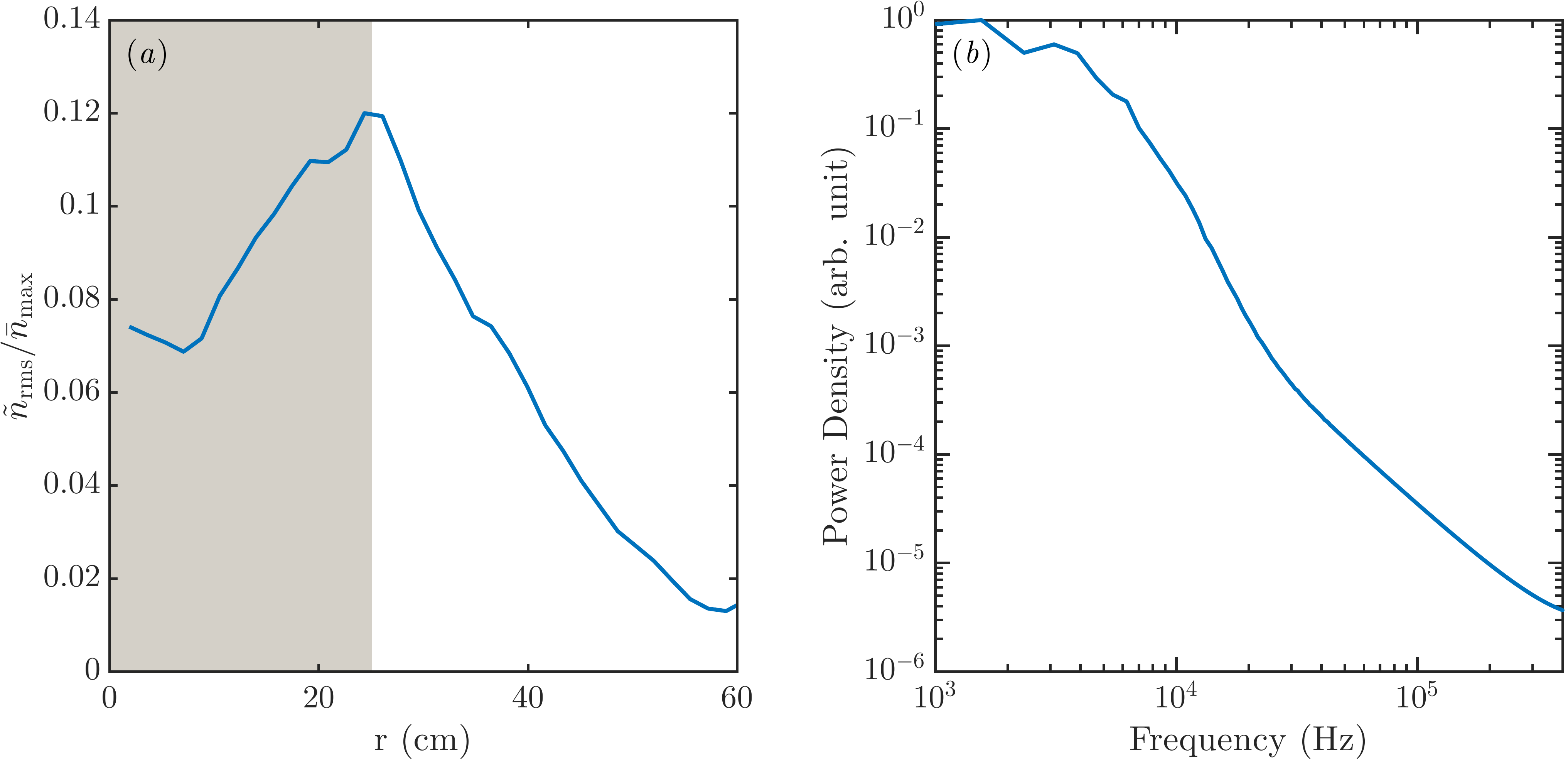}}
  \caption{Density fluctuation statistics computed from a simulation from a 5D gyrokinetic simulation of a turbulent LAPD plasma.
  ($a$) shows the normalized RMS density fluctuation level (normalizing to a constant $\bar{n}_\mathrm{max} = 3.3421 \times 10^{18}$ m$^{-3}$)
  as a function of radius and ($b$) shows the density-fluctuation power spectral density. These plots are in good
  qualitative agreement with LAPD measurements \citep{Carter2009,Friedman2012}. The shaded region in $(a)$ illustrates
  the extent of the strong plasma source.}
\label{fig:rmsDensity}
\end{figure}

Electron density fluctuation profiles have also been measured in LAPD \citep{Carter2009,Friedman2012}.
We define the density fluctuation as $\tilde{n}_e(x,y,z,t) = n_e(x,y,z,t) - \bar{n}_e(x,y,z)$,
where $\bar{n}(x,y,z)$ is computed by averaging the electron density using a 1 $\mu$s sampling interval
over a period of 1 ms.
The density fluctuation level is normalized to the peak amplitude of $\bar{n}_e$ at $r = 0$ \citep[as done in][]{Friedman2012} and binned by radius
in to calculate the RMS density fluctuation level as a function of radius,
which is shown in figure \ref{fig:rmsDensity}$(a)$.
Figure \ref{fig:rmsDensity}$(b)$ shows the power spectral density of electron density fluctuations,
which is computed by averaging the power spectra at each node
in the region $25$ cm $<r<30$ cm and $-4$ m $<z<4$ m.
Similar to measurements made on LAPD, we find that the turbulence has a broadband spectra.

\begin{figure}
  \centerline{\includegraphics[width=\textwidth]{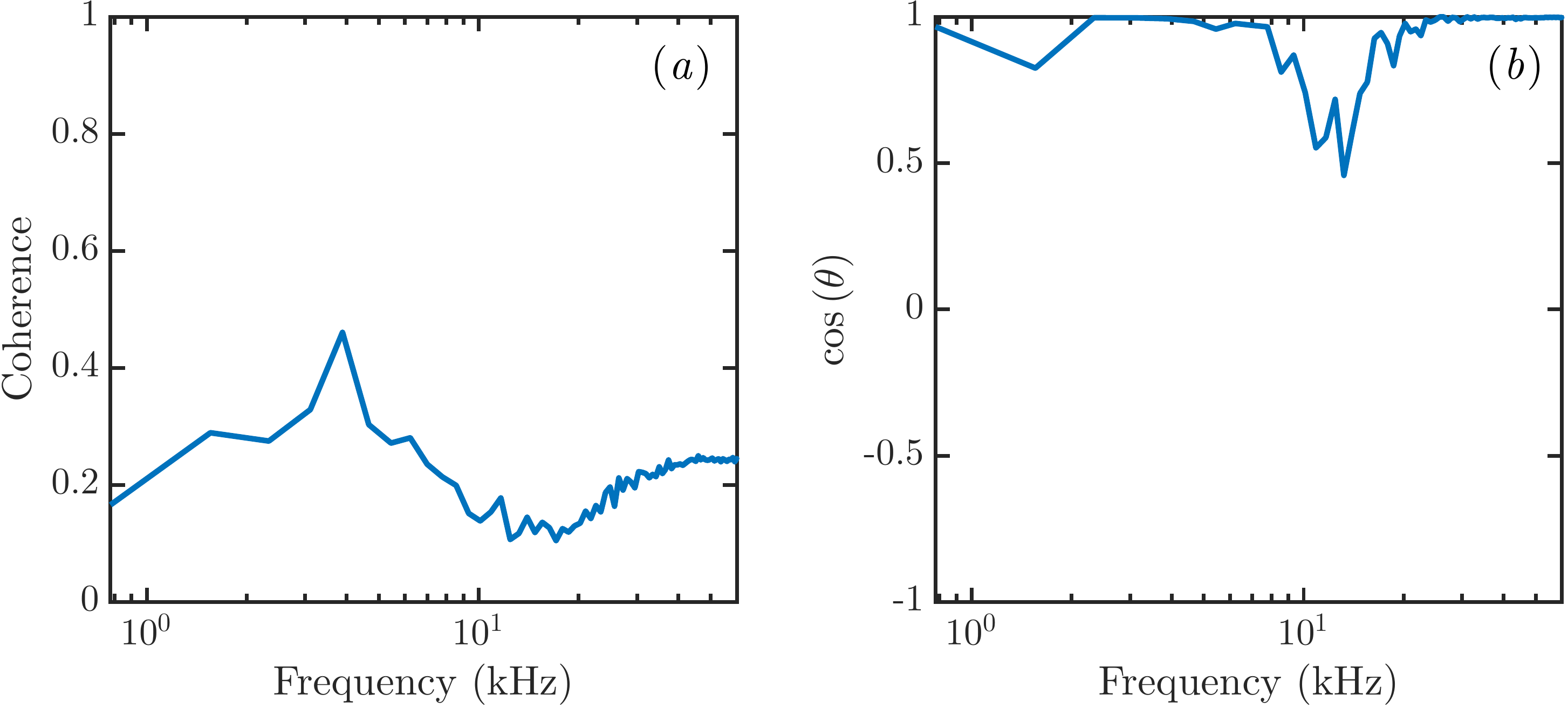}}
  \caption{Spectra of the $(a)$ coherence and $(b)$ cosine of the cross-phase between electron density
  and azimuthal electric field fluctuations. }
\label{fig:phase-and-coherence}
\end{figure}

The coherence spectrum and cross-phase spectrum between electron density fluctuations $\tilde{n}_e$ and 
azimuthal electric field fluctuations $\tilde{E}_\theta$ have also been of interest in previous 
LAPD studies for their potential role in turbulent-particle-flux suppression by applied flow shear
\citep{Carter2009,Schaffner2012,Schaffner2013}.
The cross-power spectrum $P_{nE}(f)$ is first computed at each node as:
\begin{equation}
P_{nE}(f) = \hat{n}_e^* \hat{E}_\theta,
\end{equation}
where $\hat{n}_e(\boldsymbol{x},f)$ and $\hat{E}_\theta(\boldsymbol{x},f)$ are the Fourier transforms of the time-series of
$\tilde{E}_\theta (\boldsymbol{x},t)$ and $\tilde{n}_e(\boldsymbol{x},t)$.
The cross-power spectrum is then spatially averaged, and the cross-phase is computed as
\begin{equation}
\theta(f) = \mathrm{Im} \log \left( \langle P_{nE}(f) \rangle \right),
\end{equation}
where $\langle \dots \rangle$ denotes a spatial average in the region $25$ cm $<r<30$ cm and $-4$ m $<z<4$ m.
The coherence spectrum is defined as \citep{Powers1974}
\begin{equation}
| \gamma_{nE}(f) | = \frac{| \langle P_{nE}(f) \rangle | }{ \langle P_{nn}(f) \rangle^{1/2} \langle P_{EE}(f) \rangle^{1/2} },
\end{equation}
where $P_{nn}$ and $P_{EE}$ are the real-valued power spectra of $\tilde{n}_e$ and $\tilde{E}_\theta$, respectively.
Figure \ref{fig:phase-and-coherence} shows the coherence and cross-phase spectra computed from our simulation, which
are similar to the spectra measured in LAPD \citep[see][p. 7]{Carter2009} at frequencies below 10 kHz,
where the fluctuation levels are the strongest as indicated in figure \ref{fig:rmsDensity}$(b)$.

\begin{figure} \centerline{\includegraphics[width=\textwidth]{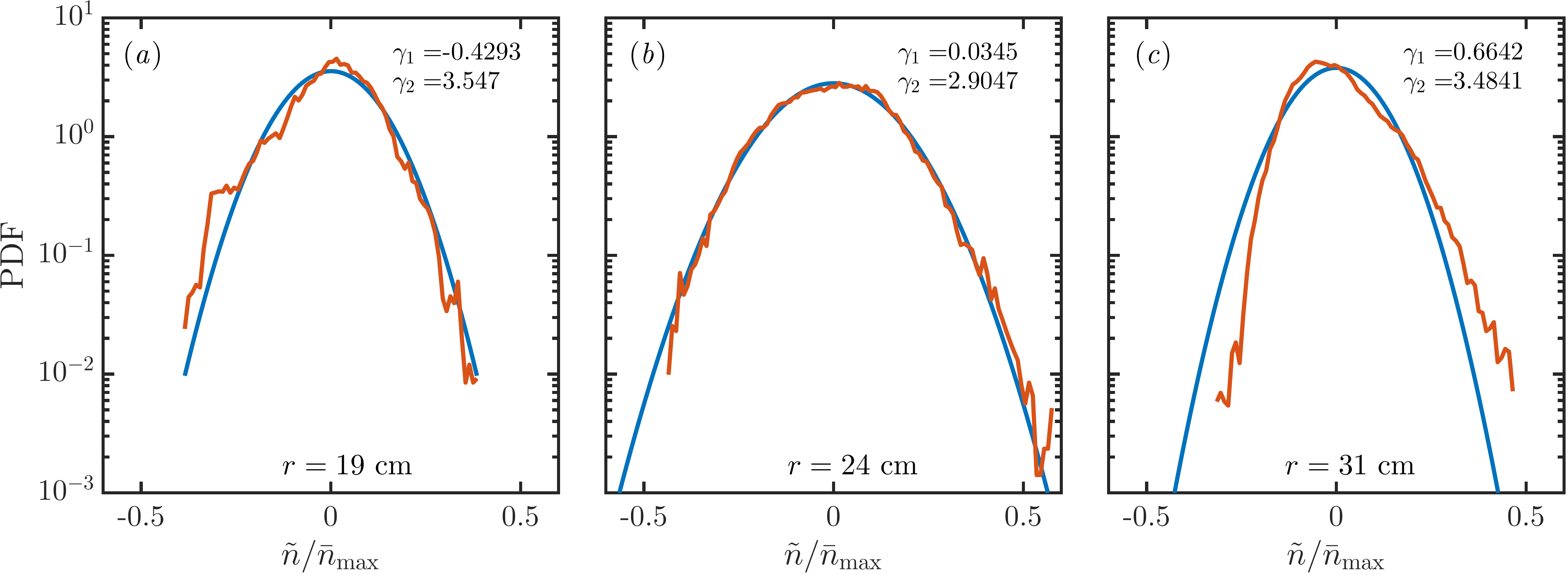}}
  \caption{Density fluctuation amplitude PDF (in red and normalized to $\bar{n}_\mathrm{max} = 3.3421 \times 10^{18}$ m$^{-3}$) at
  three radial locations in the region $-4$ m $<z<4$ m: $(a)$ slightly inside the strong-source region at $r = 19$ cm,
  $(b)$ at the location of peak fluctuation amplitude at $r=24$ cm, and $(c)$ in the weak-source region at $r = 31$ cm.
  Gaussian PDFs are shown in blue for comparison. Also indicated on each plot is the skewness $\gamma_1 = E[\tilde{n}_e^3]/\sigma^3$
  and the kurtosis $\gamma_2 = E[\tilde{n}_e^4]/\sigma^4$, where $\sigma$ is the standard deviation of $\tilde{n}_e$ and $E[\dots]$
  denotes the expected value.}
\label{fig:pdfDensity}
\end{figure}

The probability density function (PDF) of density fluctuations in LAPD has also been of interest.
\citet{Carter2006} focused on the intermittency of the density fluctuation PDF measured at various radial locations.
As shown in figure \ref{fig:pdfDensity}, we observe similar trends in our simulations, where we have
measured the PDF at three radial locations (using $\Delta r = 0.5$ cm wide radial intervals) in the region $-4$ m $<z<4$ m.
We find a negatively skewed PDF inside the strong-source region,
a symmetric and Gaussian PDF at the location of peak fluctuation amplitude, and
a positively skewed PDF in the weak-source region.
The PDF in the weak-source region has a particularly strong enhancement of large-amplitude
positive-density-fluctuation events.

\section{Conclusions \label{sec:conclusion}}
We have presented results from the first 3D2V gyrokinetic continuum simulations of turbulence in an open-field-line plasma.
The simulations were performed using a version of the Gkeyll code that employs an energy-conserving discontinuous Galerkin algorithm.
We found it important to include self-species collisions in the electrons to avoid driving
high-frequency instabilities in our simulations.
Our gyrokinetic simulations are in good qualitative 
agreement with previous Braginskii fluid simulations of LAPD and with experimental data.

We use sheath model boundary conditions for electrons that are a kinetic
extension of the sheath model used in past fluid simulations,
which allows self-consistent currents
to fluctuate in and out of the wall.
In this approach, the sheath potential is determined from the gyrokinetic Poisson
equation (analogous to how the vorticity equation is used in the fluid approach
of \citet{Rogers2010}).
The ion boundary conditions used at present are the same as for the logical sheath model,
in which ions flow out at whatever velocity they have
been accelerated to at the sheath edge.
This works well for the time period of this LAPD simulation.
As discussed in \S\ref{sec:sheath}, future work is planned to consider improved models of
a kinetic sheath, including the role of rarefaction dynamics near the sheath that may
modify the outflowing distribution function and the effective outflow Mach number.

A number of possible modifications to the simulations could allow closer quantitative modeling of the LAPD experiment.
In the real LAPD experiment, a cathode-anode discharge emits an energetic 40-60 eV electron beam that ionizes
the background gas along the length of the device
\citep{Gekelman2016,Carter2009},
creating the bulk plasma source that we have directly modelled in our simulations.
At present, we are ignoring the current from these energetic electrons
and modelling the anode as a regular conducting end plate.
Because the anode in the actual device is a high transparency mesh, there is finite pressure
on the other side of the anode from the main plasma that
can act to slow down ion outflows and thus relax the Bohm sheath criterion.
Since our simulations are gyrokinetic, future work could include the non-Maxwellian high-energy electrons and a model of the ionization process 
instead of using explicit source terms.
We have also performed simulations of turbulence suppression experiments \citep{Schaffner2012,Schaffner2013}
on LAPD using a biasable limiter to control flow shear, and these results will be presented in a future publication.
Future work will also investigate the mechanism driving the turbulence observed in our simulations by
analysing the energy dynamics of the system \citep{Friedman2012,Friedman2013}.

We plan several improvements to our numerical algorithms.
The time step restriction in our LAPD simulations is currently set by the electron-electron collision frequency.
A Super-time-stepping method, such as the Runge-Kutta-Legendre method \citep{Meyer2014}, or implicit method
could significantly alleviate this restriction.
The use of non-polynomial basis functions \citep{Yuan2006} for efficient velocity-space discretization
is expected to reduce the computational cost of these simulations (by allowing for a coarser velocity-space grid)
and to preserve the positivity of the distribution function.
Future studies will also implement the full nonlinear ion polarization density in gyrokinetic Poisson equation (\ref{eq:gkp}),
which is related to removing the Boussinesq approximation in fluid models \citep{Dudson2015,Halpern2016}.

Although the results presented here are a major milestone in our efforts towards developing a gyrokinetic continuum
code to study tokamak edge turbulence, many physical effects remain to be added to the code, such as
realistic tokamak magnetic geometry (including both open and closed-magnetic-field-line regions, a separatrix, and the X-point),
full Landau collisions, finite-Larmor-radius effects, electromagnetic effects,
and interactions with neutrals and other atomics physics.

\begin{acknowledgments}
We thank P.\,Ricci for suggesting LAPD as a test problem, T. Carter and G. Rossi for useful discussions about LAPD,
and N.\,Mandell for building the Gkeyll code on various clusters.
We also thank B.\,Friedman, J.\,Loizu, P.\,Ricci, B.\,Rogers, and M.\,Dorf
for useful discussions about various aspects of these kinds of simulations and plasma sheaths.
This work was funded by the U.S. Department of Energy under Contract No. DE-AC02-09CH11466, through
the Max-Planck/Princeton Center for Plasma Physics and the Princeton Plasma Physics Laboratory.
Initial development used the Edison system at the National Energy Research Scientific Computing Center,
a DOE Office of Science User Facility supported by the Office of Science of the
U.S. Department of Energy under Contract No. DE-AC02-05CH11231.
G.\,W.\,H. and A.\,H. were supported in part by the SciDAC Center for the Study of Plasma Microturbulence.
A.\,H. was also supported in part by the Laboratory Directed Research and Development program.
Simulations reported in this paper were performed at the TIGRESS high performance computer centre at Princeton University, which is jointly
supported by the Princeton Institute for Computational Science and Engineering and the Princeton
University Office of Information Technology's Research Computing department.

\end{acknowledgments}
\bibliographystyle{jpp}
\bibliography{shi-lapd-2017}

\end{document}